\documentclass[12pt,preprint]{aastex}

\shorttitle{Observational constraints on cosmology from
		modified Friedmann equation}
\shortauthors{Zhu, Fujimoto, \& He}

\begin{document}

\title{
	Observational constraints on cosmology from \\
		modified Friedmann equation
	}

\author{Zong-Hong Zhu}
\affil{National Astronomical Observatory, 
	2-21-1, Osawa, Mitaka, Tokyo 181-8588, Japan}
\email{zong-hong.zhu@nao.ac.jp}
 
	\and

\author{Masa-Katsu Fujimoto}
\affil{National Astronomical Observatory, 
        2-21-1, Osawa, Mitaka, Tokyo 181-8588, Japan}
\email{fujimoto.masa-katsu@nao.ac.jp}

        \and

\author{Xiang-Tao He}
\affil{Department of Astronomy, Beijing Normal University,
	Beijing 100875, China}
\email{xthe@bnu.edu.cn}

\begin{abstract}
   Recent measurements of type Ia supernovae as well as other concordant
     observations suggest that the expansion of our universe is accelerating.
   A dark energy component has usually been invoked as the most feasible 
     mechanism for the acceleration.
   However, the effects arising from possible extra dimensions can mimic
     well the role of a dark energy through a modified Friedmann equation.
   In this work, we investigate some observational constraints on a scenario
     in which this modification is given by 
     $H^2 = {8\pi G \over 3} (\rho +C\rho^n)$.
   We mainly focus our attention on the constraints from recent measurements 
     of the dimensionless coordinate distances to type Ia supernovae and
     Fanaroff-Riley type IIb radio galaxies compiled by
	Daly and Djorgovski (2003)   
     and the X-ray gas mass fractions in clusters of galaxies published by
	Allen et al. (2002,2003). 
   We obtain the confidence region on the power index $n$ of the modificative
     term and the density parameter $\Omega_m$ of the universe from
     a combined analysis of these databases.
   It is found that 
     $n=0.06^{+0.22}_{-0.18}$ and $\Omega_m=0.30^{+0.02}_{-0.02}$,
     at the 95.4\% confidence level,
     which is consistent within the errors with the standard $\Lambda$CDM model.
   These parameter ranges give a universe whose expansion swithes from 
     deceleration to acceleration at a redshift between 0.52 to 0.73.
\end{abstract}


%

\section{Introduction}

The Hubble expansion, the Cosmic Microwave Background Radiation(CMBR),
  the primordial Big Bang Nucleosynthesis and the structure formation
  are the four pillars of the standard Big Bang cosmology.
Recent years, it seems that all these cornerstoness combined to point out
  that the expansion of the universe is speeding up rather than slowing down
  (for a recent review see Peebles and Ratra 2003).
The main evidence comes from the recent well known distance measurements of
  some distant Type Ia supernovae
  (Perlmutter et al. 1998, 1999; Riess et al. 1998, 2001).
Possible explanations for such an acceleration include:
  a cosmological constant $\Lambda$ 
	(Weinberg 1989; 
	Carroll et al. 1992; 
	Krauss and Turner 1995; 
	Ostriker and Steinhardt 1995;
	Chiba and Yoshii 1999),
  a decaying vacuum energy density or a time varying $\Lambda$-term
	(Ozer and Taha 1987; 
	Vishwakarma 2001; 
	Alcaniz and Maia 2003;
	Jain, Dev and Alcaniz 2003),
  an evolving scalar field 
	(referred to by some as quintessence: 
	Ratra and Peebles 1988; 
	Caldwell et al. 1998;
	Wang and Lovelace 2001;
	Li, Hao and Liu 2002;
	Weller and Albrech 2002;
	Li et al. 2002a,b;
	Chen and Ratra 2003;
	Mukherjee et al. 2003),
  the phantom energy, in which the sum of the pressure and energy
    density is negative
        (Caldwell 2002;
        Hao and Li 2003a,b;
        Dabrowski et al. 2003),
  the so-called ``X-matter" 
	(Turner and White 1997; 
	Zhu 1998,2000;
	Waga and Miceli 1999;
	Podariu and Ratra 2001;
	Zhu, Fujimoto and Tatsumi 2001; 
	Sereno 2002;
	Alcaniz, Lima and Cunha 2003; 
	Lima, Cunha and Alcaniz  2003),
  and the Chaplygin gas 
	(Kamenshchik et al. 2001; 
	Bento et al. 2002; 
	Alam et al. 2003; 
	Alcaniz, Jain and Dev 2003; 
	Dev, Alcaniz and Jain 2003; 
	Silva and Bertolami 2003;
	Makler et al. 2003).
All the above mechanisms for accelerating are obtained by introducing a new
  hypothetical energy component with negative pressure -- the dark energy.

On the hand, many models have appeared that make use of the very ideas of 
  branes and extra dimensions to obtain an accelerating universe 
  	(Randall and Sundrum 1999a,b; 
	Deffayet et al. 2002;
	Avelino and Martins 2002; 
	Alcaniz, Jain and Dev 2002;
	Jain, Dev and Alcaniz 2002).
The basic idea behind these braneworld cosmologies is that our observable
  universe might be a surface or a brane embedded in a higher dimensional bulk 
  spacetime in which gravity could spread (Randall 2002).  
The bulk gravity see its own curvature term on the brane which accelerate the
  universe without dark energy.
Here we are concerned with the cosmological model from the modified Friedmann
  equation as follows
\begin{equation}
\label{eq:ansatz}
H^2 = {8\pi G\over 3} (\rho+C\rho^n).
\end{equation}
Freese and Lewis (2002) showed the above term proportional to $\rho^n$ (dubbed
as Cardassion term by the authors) may arise as a consequence of embedding
our observable universe as a (3+1)-dimensional brane in extra dimensions,
though an elegant and unique 5-dimensional energy-momentum tensor 
$T_{\mu\nu}$ that gives rise to equation (1) has not yet been found.
If $n<1$, the new term dominates at late times which implies a modification
of gravity at very low energy scales.
Particularly, if $n<2/3$, it gives rise to a positive acceleration.
Note that, in this scenario, although the universe is flat and accelerating,
  it contains matter (and radiation) only without any dark energy contribution.

The main goal of this paper is to set observational constraints on this
  Cardassian expansion model and check whether it is consistent with current
  cosmological data. 
We perform a combined analysis of data including the dimensionless coordinate
  distances to type Ia supernovae (SNeIa) and Fanaroff-Riley type IIb (FRIIb) 
  radio galaxies compiled by
	Daly and Djorgovski (2003)
  and the X-ray gas mass fraction in clusters of galaxies published by
	Allen et al. (2002,2003).
These results are timely and complementary to the previous constraints from
  the angular size of high-$z$ compact radio sources (Zhu and Fujimoto 2002), 
  CMBR (Sen and Sen 2003a,b),
  the SNeIa database (Zhu and Fujimoto 2003a; Wang et al. 2003; Cao 2003; 
	Szydlowski and Czaja 2003; Godlowski and Szydlowski 2003),
  large scale structures (Multamaki et al. 2003)
  and from optical gravitational lensing surveys (Dev, Alcaniz and Jain 2003).

The plan of the paper is as follows.
In the next section, we provide a brief summary of the Cardassian expansion
  scenario relevant to our work.
Constraints from dimensionless coordinate distance data of SNeIa and FRIIb 
  radio galaxies are discussed in section~3.
In section~4 we discuss the bounds imposed by X-ray gas mass fraction in
  galaxy clusters.
Finally, we present a combined analysis, our concluding remarks 
  and discussion in section~5.


\section{The Cardassian model: basic equations}

In the modified Friedmann equation Eq.(1), there are two model parameters,
  the power index $n$ and the coefficient $C$ of the Cardassian term.
Instead of $C$, it is convenient to use the redshift $z_{\rm car}$, at which
  the two terms inside the bracket of Eq.(1) are equal, as the second
  parameter of the model, i.e, 
  $C = [\rho(z_{\rm car})]^{1-n}$.
If we further ignore the radiation components in the universe that are not
  important for the cosmological tests considered in this work, we have
  $\rho = \rho_0 (1+z)^3$, $\rho (z_{\rm car}) =\rho_0 (1+z_{\rm car})^3$ and
\begin{equation}
H^2= {8\pi G\over 3} \rho_0 (1+z)^3 
	\left[ 1+ {\left( {1+z \over 1+z_{\rm car}} \right)^{3(n-1)}} \right]
\end{equation}
  where $\rho_0$ is the present matter density of the universe.
Hence Evaluating the Hubble parameter today gives
$
H_0^2 = {8 \pi G\over 3 } \rho_0 [1+(1+z_{\rm car})^{3(1-n)}] .
$
Conventionally, the critical density of the universe is 
  $\rho_c=3H_0^2/8\pi G = 1.8791 \times 10^{-29}h^{2}{\rm g~cm}^{-3}$, where
  $h$ is the present day Hubble constant in units of 
  100 ${\rm{km~s^{-1}Mpc^{-1}}}$, and the present matter density of the 
  universe is written in terms of $\rho_c$ as $\rho_0 = \Omega_m \rho_c$,
  where $\Omega_m$ is the standard matter density parameter.
Now in Cardassian model, matter alone makes the universe flat, which means that
  $\rho_0 = \rho_{c,{\rm car}}$ and $\rho_{c,{\rm car}}$ is the critical
  density of the universe in Cardassian expansion model (Freese and Lewis 2002)
\begin{equation}
\rho_0 = \rho_{c,{\rm car}} = 
			{3 H_0^2 \over 8 \pi G [1+(1+z_{\rm car})^{3(1-n)}]} .
\end{equation}
and the standard matter density parameter of the universe is 
  $\Omega_m = [1+(1+z_{\rm car})^{3(1-n)}]^{-1}$
  (note that $\Omega_{m,\rm car} \equiv \rho_0/\rho_{c,{\rm car}} =1$,
	as expected for a flat universe).
As it shown in equation (3), for some combinations of the parameters 
  $n$ and $z_{\rm car}$, the critical density of the Cardassion model 
  can be much lower than the one of the standard Friedmann model.
In other words, in the context of Cardassian model, it is 
  possible to make the dynamical estimates of the quantity of matter that 
  consistently point to $\rho_0 \simeq (0.2-0.4)\rho_c$ compatible with the 
  observational evidence for a flat universe from CMB observations and the 
  flatness prediction made by inflationary cosmology without any dark 
  energy component (see Freese and Lewis 2002 for more details).

Now we evaluate the dimensionless coordinate distance, $y(z)$,  
  the angular diameter distance, $D^A (z)$, and
  the luminosity distance, $D^L (z)$, as a function of redshift
  $z$ as well as the parameters of the Cardassian model.
The three distances are simply related to each other by
  $D^L = (1+z)^2 D^A = (c/H_0) (1+z) y(z)$. 
Following the notation of Peebles (1993), we define the redshift
  dependence of the Hubble paramter $H$ as $H(z) = H_0 E(z)$.
Parametrizing the model as $(\Omega_m, n)$, we get $E$ function as
\begin{equation}
\label{eq:newE}
E^2(z; \Omega_m, n) = \Omega_m (1+z)^3 + (1-\Omega_m) (1+z)^{3n}
\end{equation}
We note that the quintessence models with a constant equation of
  state ($p=\omega\rho$) for the dark energy component give rise
  to the same $H(z)$ presented here.
One can make the following identification: $\omega = n -1$, such
  that $n=1$ correponds to a $\Lambda$CDM model.
As far as any observation that involves only $H(z)$, the two
  models predict the same effects on the observation (Freese 2003).
It is straightforward to show that the distances are given by
\begin{equation}
\label{eq:DA}
D^L(z; H_0, \Omega_m, n) = (1+z)^2 \cdot D^A(z; H_0, \Omega_m, n)
                         = {c \over H_0 } (1+z) \cdot y(z; \Omega_m, n) 
                         = {c \over H_0 } (1+z) \cdot \int_{0}^{z} {dz^{\prime}
				\over E(z^{\prime}; \Omega_m, n) }
\end{equation}
%


\section{Constraints from the dimensionless coordinate distance data}

Recently, Daly and Djorgovski (2003) compiled a large database of the
  dimensionless coordinate distance measurements estimated from the
  observations of SNeIa and FR IIb radio galaxies.
The author used the database to derive the expansion rate of the universe 
  as a function of redshift, $E(z)$, and the acceleration rate of the 
  universe as a function of redshift, $q(z)$ (Daly and Djorgovski 2003).
We use this sample to give an observational constraint on the 
  model parameters, $n$ and $\Omega_m$.

The SNeIa measurements include 
  the 54 supernovae in the ``primary fit C'' used by Perlmutter et al. (1999), 
  the 37 supernovae published by Riess et al. (1998), and
  the so far highest redshift supernova 1997ff presented by Reiss et al. (2001).
The standard procedures of Perlmutter et al. (1999) and Riess et al. (1998)
  were used to determine the dimensionless coordinate distances to the 
  supernovae.
The apparent bolometric magnitude $m(z)$ of a standard candle with absolute
  bolometric magnitude $M$ is related to the luminosity distance $D^L$ by
  $m=M+5\log D^L + 25$.
Then using the relation of equation (5) 
	the B-band magnitude-redshift relation can be written as
\begin{equation}
\label{eq:mB}
m_B = {\cal M}_B + 5 \log [c(1+z) \cdot y(z)]
\end{equation}
where ${\cal M}_B \equiv M_B - 5 \log H_0 + 25$ is the 
  ``Hubble-constant-free'' $B$-band absolute magnitude at maximum of a
  type Ia supernova (SNIa).
Then the above relation is used to determine $y(z)$ for each SNIa.
There are 14 SNeIa that are present in both the Perlmutter et al. (1999)
  and Riess et al. (1998) samples, for which we will use the average values
  of $y$ with appropriate error bars (see Table 4 of Daly and Djorgovski 2003).
Therefore we totally have 78 SNIa data points that are shown as
  solid circles in Figures 1 and 2.

%
%

The use of FRIIb radio galaxies to determine the angular size distance 
  or dimensionless coordinate distance of them at different redshifts 
  was first proposed by Daly (1994) (see also Guerra, Daly, and Wan 2000; 
  Daly and Guerra 2002; Podariu et al. 2003; Daly and Djorgovski 2003). 
This method consists in a comparison of two independent measurements of the 
  average size of the lobe-lobe separation of FR IIb sources, namely, 
  the mean size $<D>$ of the full population of radio galaxies at similar 
  redshift and the source average (over its entire life) size $D_*$, that is
  determined via a physical model describing the evolution of the sources.
The basic idea is that $<D>$ must track the value of $D_*$, such that the 
  ratio $R_*=<D>/D_*$ is assumed to be a constant:
  $R_* (\beta, y(z)) = \kappa~.$, where $\beta$ is one parameter entering
  into the ratio $R_*$.
$y(z)$ can be determined using an iterative technique, as described in detail 
  by Guerra, Daly and Wan (2000) and Daly and Guerra (2002). 
We use the values of $y(z)$ for 20 FRIIb radio galaxies obtained using the best
  fit to both the radio galaxy and supernova data (see Table 1 of 
  Daly and Djorgovski 2003), that are shown as empty squares in Figure 1
  and 2.
The best fit values of $\kappa$ and $\beta$ and their error bars are included 
  in Table 2 of Daly and Djorgovski (2003), i.e.,
  $\kappa = 8.81 \pm 0.05$ and
  $\beta = 1.75 \pm 0.04$.
In determining the error bar on $y(z)$, the uncertainties of $\kappa$, and 
  $\beta$ have been included (Daly and Djorgovski 2003).
It is very important to consider whether significant covariance exists between
  the different parameters determined by the fit (Daly and Guerra 2002).
In the framework of quintessence model, Daly and Guerra (2002) have estimated
  the likelihood contours in the $\beta-\omega$ plane and the $\beta-\Omega_m$
  plane respectively (see Figure 3 and 4 of their paper).
As was pointed out in section 2, the Cardassion scenario is equivalent to
  quintessence model by identifying $\omega=n-1$.
Therefore their results are exactly appropriate to our parameters $n$ and 
  $\Omega_m$: i.e., there is no covariance between $\beta$ and $n$ ($\Omega_m$).
 
We determine the model parameters $n$ and $\Omega_m$ through 
  a $\chi^{2}$ minimization method.
The range of $n$ spans the interval [-1,1] in steps of 0.01, while the
  range of $\Omega_m$ spans the interval [0, 1] also in steps of 0.01.
\begin{equation}
\label{eq:chi2}
\chi^{2}(n, \Omega_m) =
  \sum_{i=1}^{98}{\frac{\left[y(z_i;n, \Omega_m) - 
   y_{{\rm o}i}\right]^{2}}{\sigma_i^{2}}},
\end{equation}
where $y(z_i;n, \Omega_m)$ refers to the theoretical prediction from 
  equation~(5), $y_{{\rm o}i}$ is the observed dimensionless coordinate
  distances of SNeIa and FRIIb radio galaxies, and $\sigma_i$ is the 
  uncertainty ($i$ refers to the $i$th data point, with totally 98 data).
The summation is over all of the observational data points.

The results of our analysis for the Cardassian expansion model are displayed
  in Figure 3.
We show 68.3\% and 95.4\% confidence level contours in the ($\Omega_m$,$n$)
  plane using the lower shaded and the lower plus darker shaded areas
  respectively.
The best fit happans at $\Omega_m=0.38$ and $n=-0.20$.
It is clear from the figure, that the dimensionless coordinate distance test
  alone weakly constrains the Cardassian expansion model.
Only models with $\Omega_m > 0.60$ and $n > 0.54$ are excluded at the
  $95.4\%$ confidence level. 
However, this already strongly suggest an accelerating universe
  (because $n < 0.54 < 2/3$).
Moreover, as we shall see in Sec.5, when we combine this test with the X-ray
  gas mass fraction test, we could get very stringent constraints on the
  Cardassian scenario. 

Recently, Knop et al. (2003) investigated in detail the effects of various
  systematic errors of SNeIa on the cosmological measurements (section 5
  of their paper).
Their main results were summarized in Table 9 (Knop et al. 2003), for example,
  the differences in lightcurve fitting methods can change the flat universe
  value of $\Omega_m$ by $\sim 0.03$ and $\omega$ ($=n-1$) by 0.02.
Other systematic errors considered include non-type Ia SN contamination,
  Malmquist bias, $K$-corrections, SN colors, dust extinction and gravitational
  lensing etc.
All identified systematic errors together give rise to 
  $\Delta \Omega_m = 0.04$ and $\Delta \omega$ ($=\Delta n$) = 0.09,
  which are smaller than than the current statistical uncertainties of SNeIa 
  (Knop et al. 2003).

%
%


\section{Constraints from the galaxy clusters X-ray data}

Clusters of galaxies are the largest virialized systems in the universe,
  and their masses can be estimated by X-ray and  optical observations,
  as well as gravitational lensing measurements.
A comparison of the gas mass fraction, 
  $f_{\rm gas} = M_{\rm gas} / M_{\rm tot}$,
  as inferred from X-ray observations of clusters of galaxies to the cosmic
  baryon fraction can provide a direct constraint on the density parameter
  of the universe $\Omega_m$ (White et. al. 1993).
Moreover, assuming the gas mass fraction is constant in cosmic time,
  Sasaki (1996) show that the $f_{\rm gas}$ measurements of clusters of galaxies
  at different redshifts also provide an efficient way to constrain other
  cosmological parameters decribing the geometry of the universe.
This is based on the fact that the measured $f_{\rm gas}$ values for each
  cluster of galaxies depend on
  the assumed angular diameter distances to the sources as 
  $f_{\rm gas} \propto [D^A]^{3/2}$.
The ture, underlying cosmology should be the one which make these measured
  $f_{\rm gas}$ values to be invariant with redshift 
  (Sasaki 1996; Allen at al. 2003).

Using the {\it Chandra} observational data, Allen et al. (2002; 2003) have
  got the $f_{\rm gas}$ profiles for the 10 relaxed clusters.
Except for Abell 963, the $f_{\rm gas}$ profiles of the other 9 clusters
  appear to have converged or be close to converging with a canonical radius
  $r_{2500}$, which is defined as the radius within which the mean mass 
  density is 2500 times the critical density of the universe at the redshift
  of the cluster (Allen et al. 2002, 2003).
The gas mass fraction values of these nine clusters at $r_{2500}$ (or at the
  outermost radii studied for PKS0745-191 and Abell 478) are shown 
  in Figure 4.
We will use this database to constrain the Cardassian expansion models.
Following Allen et al. (2002), we have the model function as
\begin{equation}
f_{\rm gas}^{\rm mod}(z_i;n,\Omega_m) =
      \frac{ b \Omega_b}{\left(1+0.19{h}^{1/2}\right) \Omega_m}
  \left[{h\over 0.5}
	\frac{D^A_{\rm{SCDM}}(z_i)}{D^A_{\rm car}(z_i;n,\Omega_m)}
		\right]^{3/2}
\end{equation}
where the bias factor $b \simeq 0.93$ (Bialek et al. 2001; Allen et al. 2003)
  is a parameter motivated by gas dynamical simulations, which suggest that
  the baryon fraction in clusters is slightly depressed with respect to the
  Universe as a whole 
	(Cen and Ostriker 1994; 
	Eke, Navarro and Frenk 1998;
	Frenk et al. 1999; 
	Bialek et al. 2001).
The term $(h/0.5)^{3/2}$ represents the change in the Hubble parameter from
  the defaut value of $H_0 = 50 {\rm{km \, s^{-1} \, Mpc^{-1}}}$ and
  the ratio ${D^A_{\rm{SCDM}}(z_i)}/{D^A_{\rm{car}}(z_i;n,\Omega_m)}$ 
  accounts for the deviations of the Cardassian model from the default 
  standard cold dark matter (SCDM) cosmology.

%
%

Again, we determine the Cardassian model parameters $n$ and $\Omega_m$ through
  a $\chi^{2}$ minimization method.
We constrain $\Omega_m h^{2} = 0.0205 \pm 0.0018$ (O'Meara et al. 2001)
  and $h = 0.72 \pm 0.08$, the final result from the Hubble Key Project by
  Freedman et al. (2001).
The range of $n$ spans the interval [-1,1] in steps of 0.01, while the
  range of $\Omega_m$ spans the interval [0, 1] also in steps of 0.01.
The $\chi^2$ difference between the model function and SCDM data is then
  (Allen et al. 2003)
\begin{equation}
\label{eq:chi2}
\chi^{2}(n, \Omega_m) =    \sum_{i = 1}^{9} 
  \frac{\left[f_{\rm gas}^{\rm mod}(z_i;n,\Omega_m) - f_{{\rm gas,o}i}\right]^2}
	{\sigma_{f_{{\rm gas},i}}^2}	+
  \left[\frac{\Omega_bh^{2} - 0.0205}{0.0018}\right]^{2}   +
  \left[\frac{h - 0.72}{0.08}\right]^{2},
\end{equation}
where $f_{\rm gas}^{\rm mod}(z_i;n,\Omega_m)$ refers to equation~(8),
  $f_{{\rm gas,o}i}$ is the measured $f_{\rm gas}$ with the defaut cosmology
  (SCDM and $H_0 = 50 {\rm{km \, s^{-1} \, Mpc^{-1}}}$)
  and $\sigma_{f_{{\rm gas},i}}$ is the symmetric root-mean-square
  errors ($i$ refers to the $i$th data point, with totally 9 data).
The summation is over all of the observational data points.

The results of our analysis for the Cardassian expansion model are displayed
  in Figure 5.
We show 68.3\% and 95.4\% confidence level contours in the ($\Omega_m$,$n$)
  plane using the lower shaded and the lower plus darker shaded areas
  respectively.
The best fit happans at $\Omega_m=0.30$ and $n=0.14$.
It is clear from the figure, that although the X-ray gas mass fraction test 
  alone constrains the density parameter $\Omega_m$ very stringently,
  it weakly limits the Cardassian power index $n$.
However, when comparing Figure 4 with Figure 3, we can expect the X-ray gas
  mass fraction test to break the degeneracy presented in the dimensionless
  coordinate distance test of last section.

As Figure~5 shown, measurements of $f_{\rm gas}$ of galaxy clusters provide
  an efficient way to determine $\Omega_m$.
However the uncertainty of the bias factor $b$ will lead to a systematic
  error in this kind of analysis (Allen et al. 2003).
Because it linearly scales the X-ray mass fraction $f_{\rm gas}$ in 
  equation (8), lowering (uppering) it by $\sim 10\%$ would cause the best
  fitting value of $\Omega_m$ to reduce (increase) by a similar amount.
Another systematic uncertainty comes from the $f_{\rm gas}$ profiles of 
  galaxy clusters:  any rise in the $f_{\rm gas}$ values beyond the measuremet
  radii would cause a corresponding reduction in $\Omega_m$.

%
%


\section{Combined analysis, concluding remarks and discussion}

Now we present our combined analysis of the constraints discussed in the
  previous sections and summarize our results.
In Figure 6, we display the 68.3\%, 95.4\% and 99.7\% confidence level 
  contours in the ($\Omega_m$,$n$) plane from a combination of databases of
  dimensionless coordinate distances to SNeIa and FRIIb radio galaxies 
  and the X-ray gas mass fraction in clusters of galaxies. 
The best fit happans at $\Omega_m=0.30$ and $n=0.06$.
As it shown, although
  the two Cardassion parameters are not very sensitive to 
  the dimensionless coordinate distance data of SNeIa and FRIIb
  radio galaxies and the Cardassion power index is not sensitive to
  the X-ray gas mass fraction data of clusters,
  a combination of the two data sets gives at the 95.4\% confidence level
  that $\Omega_m=0.30^{+0.02}_{-0.02}$ and $n=0.06^{+0.22}_{-0.18}$, 
  a very stringent constraint on the Cardassion expansion scenario.

%
%

Given the two model parameters, $n$, and $\Omega_m$, 
  Zhu and Fujimoto (2003b)
  derived the redshift $z_{\rm car}$ and $z_{q=0}$ satisfying the relation
  as follows
\begin{equation}
\label{eq:zq=0}
(1+z)_{q=0} = (2-3n)^{1\over {3(1-n)}} (1+z_{\rm car}) =
\left[ (2-3n)({1\over \Omega_m} -1) \right]^{1\over {3(1-n)}} \,\,.
\end{equation}
where $z_{\rm car}$ is the redshift at which the two terms inside the bracket 
  of Eq.(1) are equal, while $z_{q=0}$ is the redshift at which the universe 
  switches from deceleration to acceleration, or in other words
  the redshift at which the deceleration parameter vanishes.
It was shown (Zhu and Fujimoto 2003b) that, 
  for every value of $\Omega_m$, there exists a value for
  the power index of the Cardassian term, $n_{\rm peak}(\Omega_m)$, satisfying
  $(2-3n_{\rm peak})^{-1} \exp\left[3(1-n_{\rm peak})/(2-3n_{\rm peak})\right] = \Omega_m^{-1} -1$
which makes the turnaround redshift $z_{q=0}$ reach the maximum value,
  $[ z_{q=0} ]_{\rm max} = \exp\left[1/ (2-3n_{\rm peak})\right] -1$.
%
%
Lower $\Omega_m$ is, higher $[ z_{q=0} ]_{\rm max}$ will be.
For the lower bound obtained here, $\Omega_m=0.28$, 
  we have $n_{\rm peak}=0.0576$ and $[ z_{q=0} ]_{\rm max}=0.73$.
In conclusion, our combined analysis result, 
  $\Omega_m=0.30^{+0.02}_{-0.02}$ and $n=0.06^{+0.22}_{-0.18}$ 
  at the 95.4\% confidence level, suggest a Cardassion expansion universe
  which swithes from deceleration to acceleration around 
    $z_{q=0} \in (0.52,0.73)$.
However, the modified term of the Friedmann equation
    would dominate at a redshift around $z_{\rm car} \in (0.25,0.55)$, 
    a little bit later than the expansion turnaround happans 
    (note that, $z_{\rm car}$ is generally not equal to $z_{q=0}$),
    which is simply due to the resulting power index $n$ is well below $1/3$
    (Zhu and Fujimoto 2003b).

Alternative cosmologies from a modified Friedmann equation may provide a
  possible mechanism for the present acceleration of the universe congruously
  suggested by various cosmological observations.
In this paper we have focused our attention on one of these scenarios, the
  so-called Cardassion expansion in which the universe is flat, matter 
  (and radiation) dominated and accelerating 
  but without any dark energy component.
We have shown that stringenet constraints on the parameters $n$ and 
  $\Omega_m$, that completely characterize the scenario, 
  can be obtained from the combination analysis of the dimensionless
  coordinate distance data of SNeIa and FRIIb radio galaxies and the X-ray 
  mass fraction data of clusters.
It is natually hopeful that, with a more general analysis such as a joint
  investigation on various cosmological observations, one could show clearly
  if this scenario constitutes a feasible alternative to other acceleration
  mechanisms.

\acknowledgements

We would like to thank 
  S. Allen for sending us their compilation of the X-ray mass fraction data and
    helpful explanation of the data,
	J. S. Alcaniz, 
	A. Dev,
	D. Jain
	and D. Tatsumi for their helpful discussion.
Our thanks go to the anonymouse referee for valuable comments and useful
  suggestions, which improved this work very much.
This work was supported by
  a Grant-in-Aid for Scientific Research on Priority Areas (No.14047219) from
  the Ministry of Education, Culture, Sports, Science and Technology.

\clearpage

\begin{figure}
\plotone{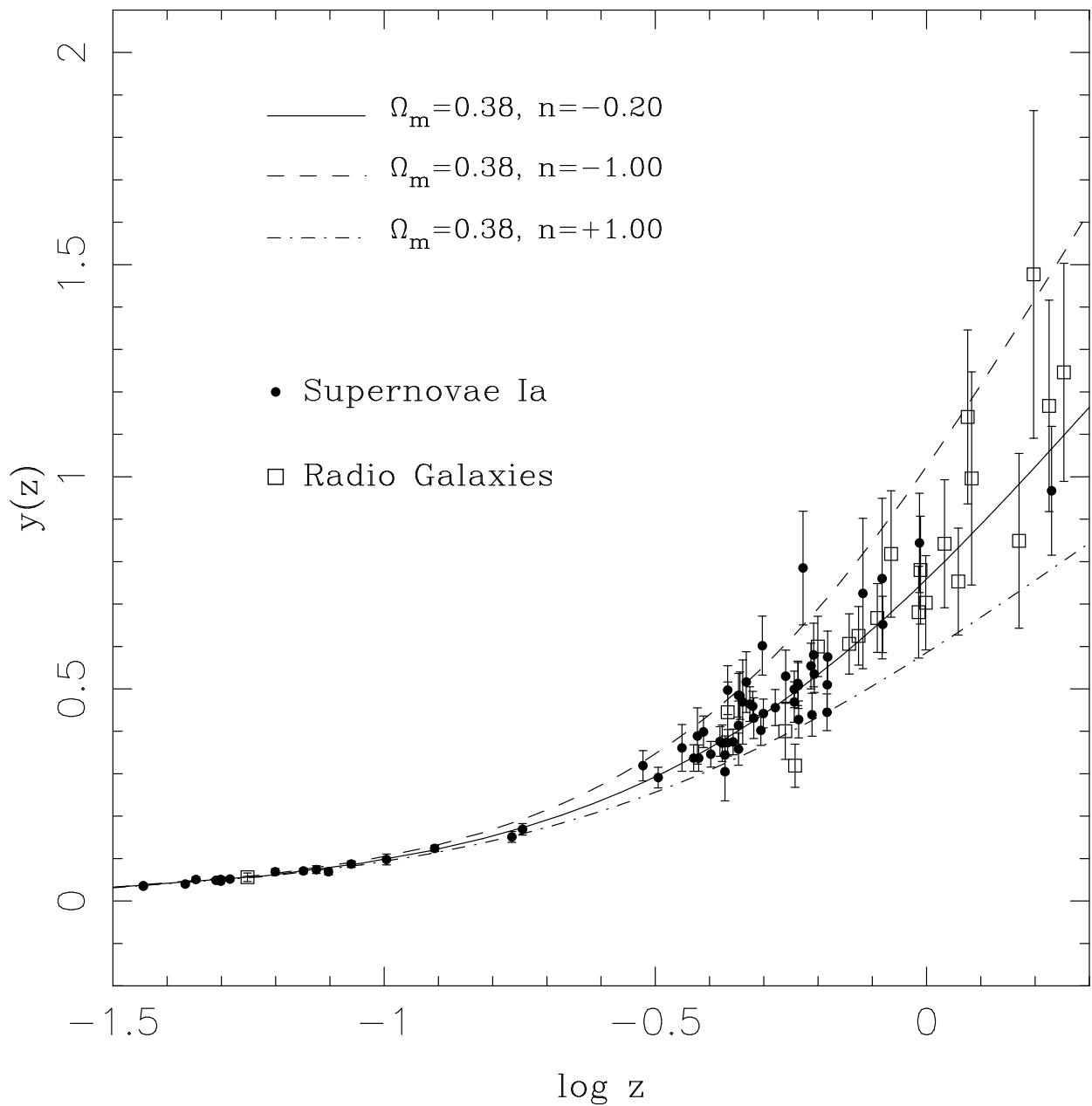}
\figcaption{Dimensionless coordinate distances $y(z)$ as a function of $\log z$
            for 78 SNeIa and 20 FRIIb radio galaxies.
            The solid circles mark the SNeIa, while the empty squares mark the
            FRIIb radio galaxies.
            The solid curve corresponds to our best fit to the total 98 data
            points with $\Omega_m=0.38, n=-0.20$.
            The database are taken from Daly and Djorgovski (2003).
	    \label{Fig_data11}
           }
\end{figure}

\clearpage

\begin{figure}
\plotone{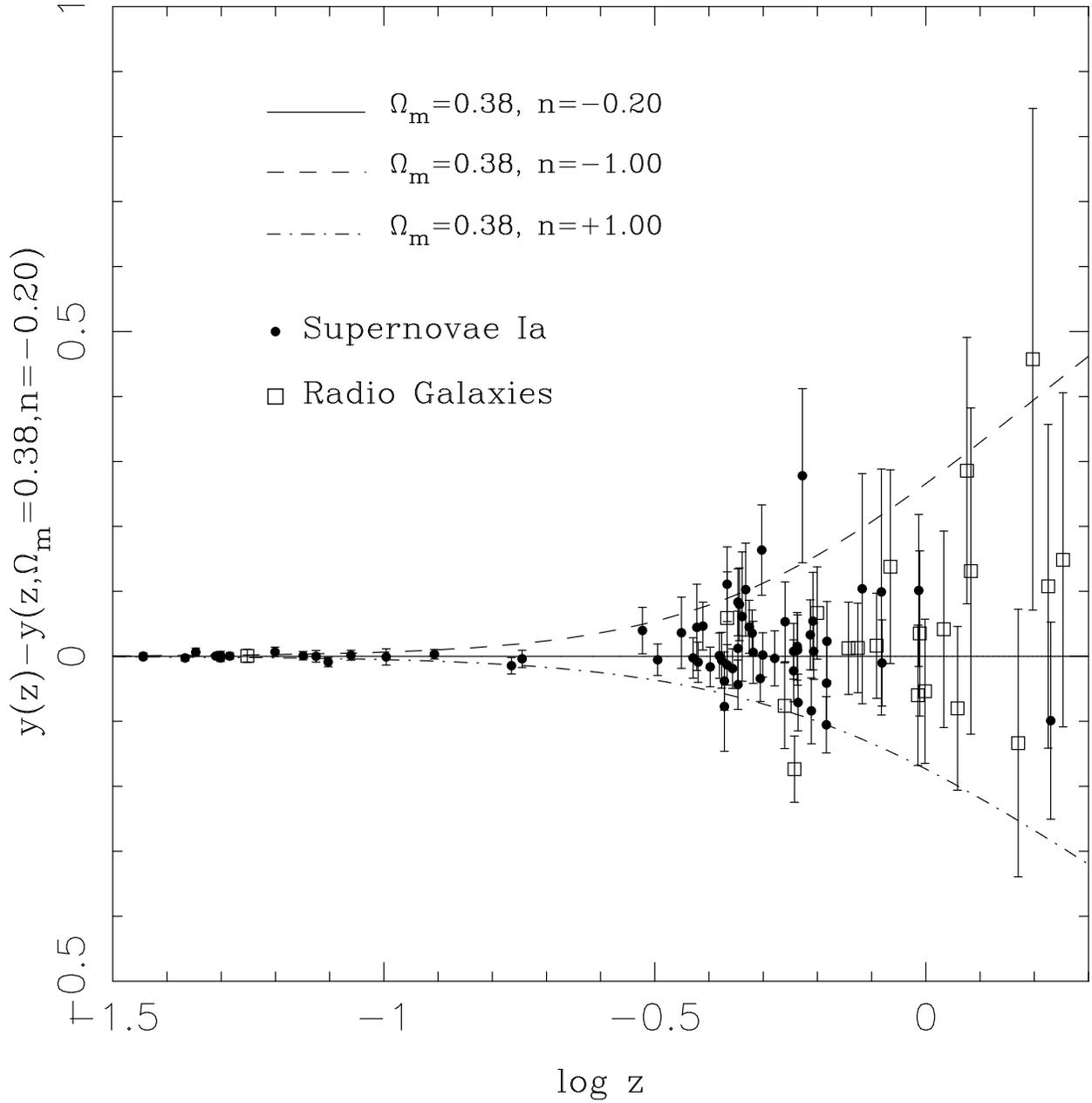}
\figcaption{The residuals of $y(z)$ relative to our best fit model
            with $\Omega_m=0.38$, and $n=-0.20$
            as a function of $\log z$ for 78 SNeIa
            (the solid circles) and 20 FRIIb radio galaxies (the empty squares).
            The data are taken from Daly and Djorgovski (2003).
	    \label{Fig_data12}
	   }
\end{figure}

\clearpage

\begin{figure}
\plotone{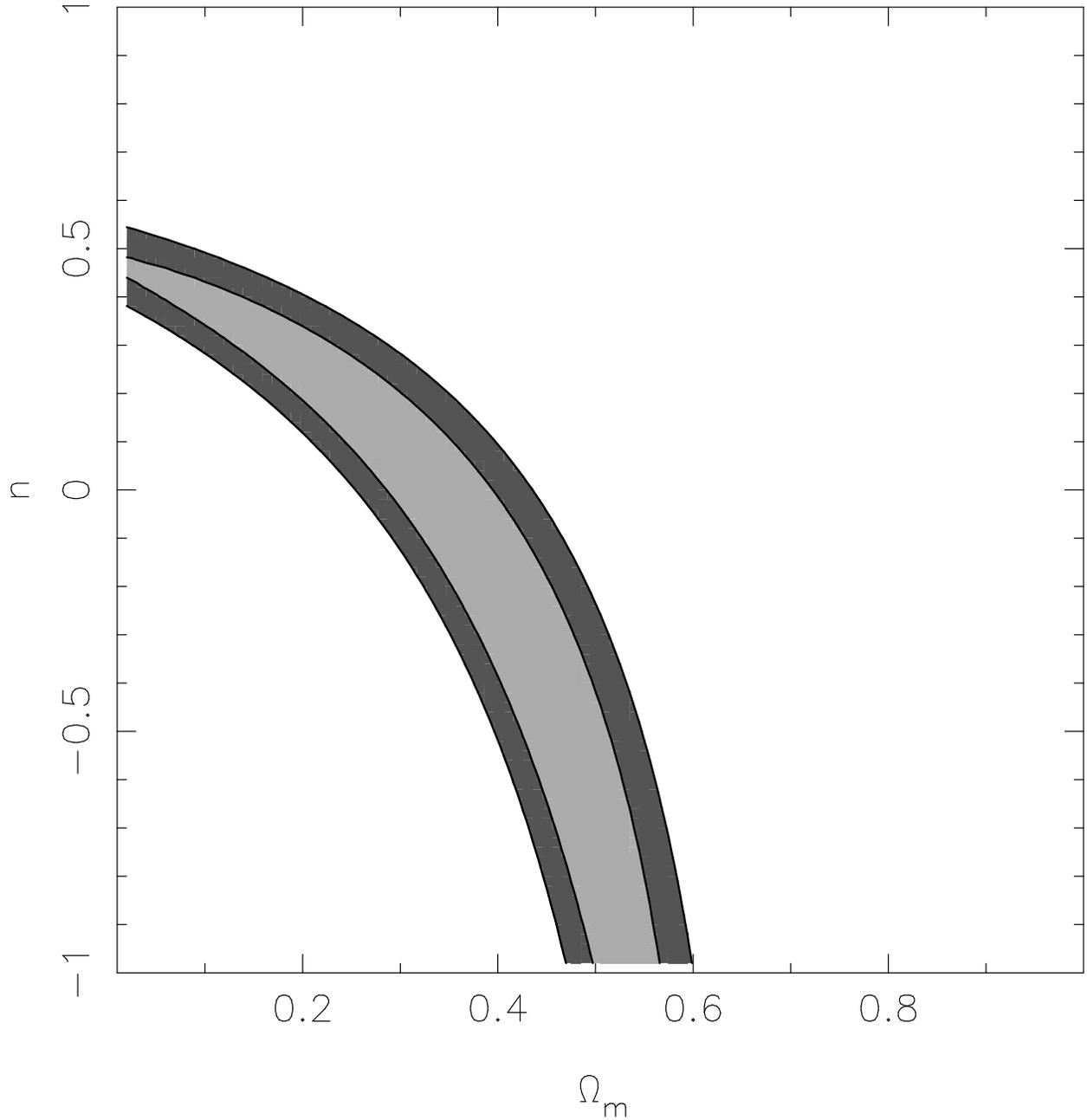}
\figcaption{Confidence region plot of the best fit to the database of the
            dimensionless coordinate distances to 78 SNeIa and 20 FRIIb
            radio galaxies compiled by Daly and Djorgovski (2003) --
            see the text for a detailed description of the method.
            The 68.3\% and 95.4\% confidence levels in the $n$--$\Omega_m$ plane
            are shown in lower shaded and lower $+$ darker shaded areas
            respectively.
	    \label{Fig_cont1}
	   }
\end{figure}

\clearpage

\begin{figure}
\plotone{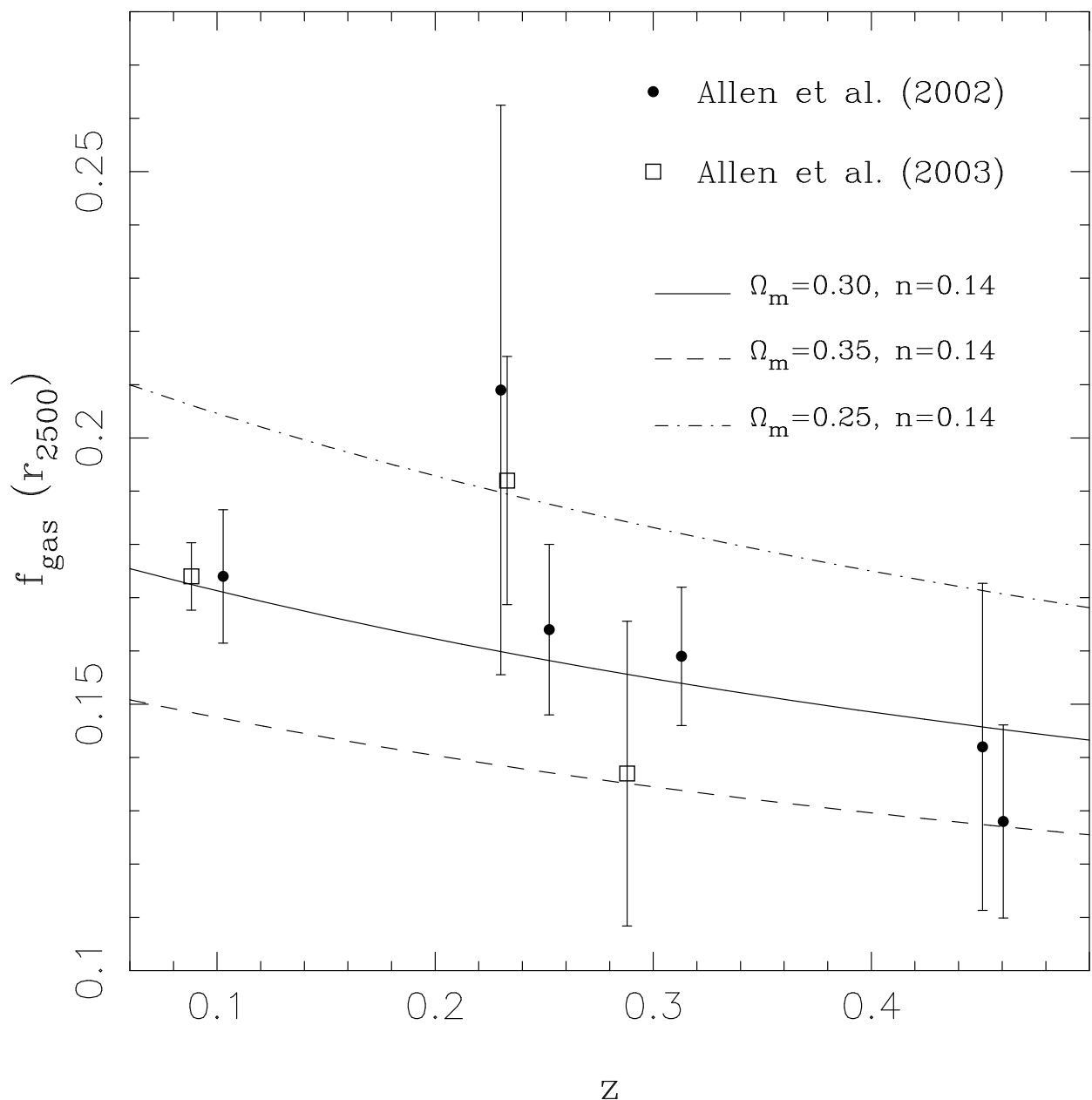}
\figcaption{The apparent redshift dependence of the $f_{\rm gas}$ measured
            at $r_{2500}$ for 9 clusters of galaxies with convergent
            $f_{\rm gas}$ profiles.
            The error bars are the symmetric root-mean-square $1\sigma$ errors.
            The solid circles mark the six clusters studied by
            Allen et al.(2002), while the empty squares mark the other
            three clusters published by Allen et al. (2003).
            The solid curve corresponds our best fit to the Cardassian model
            with $\Omega_m = 0.30$, and $n =0 .14$.
	    \label{Fig_data2}
	   }
\end{figure}

\clearpage

\begin{figure}
\plotone{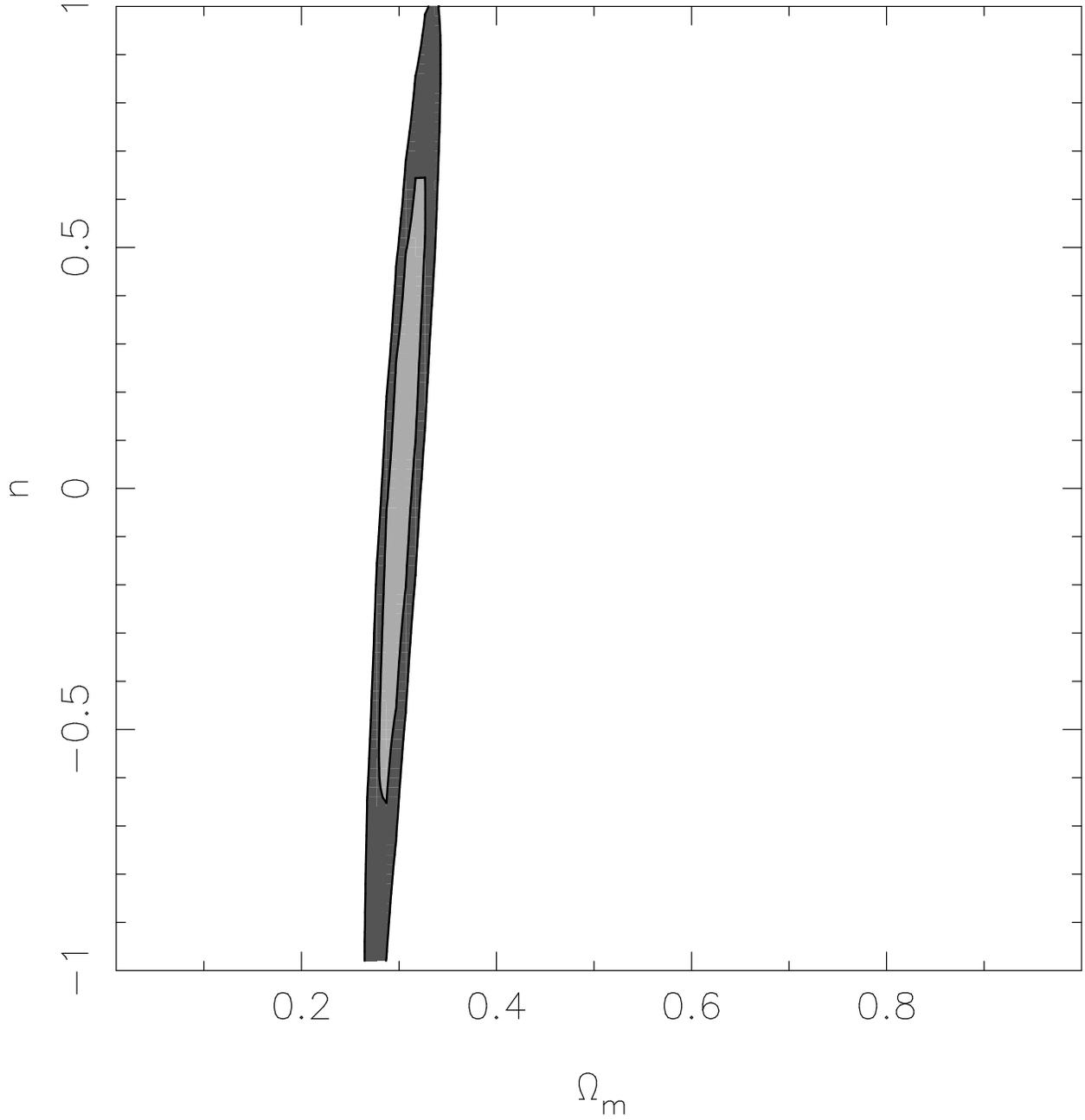}
\figcaption{Confidence region plot of the best fit to the $f_{\rm gas}$ of
            9 clusters published by Allen et al. (2002,2003) --
            see the text for a detailed description of the method.
            The 68.3\% and 95.4\% confidence levels in the $n$--$\Omega_m$ plane
            are shown in lower shaded and lower $+$ darker shaded areas
            respectively.
	    \label{Fig_cont2}
           }
\end{figure}

\clearpage

\begin{figure}
\plotone{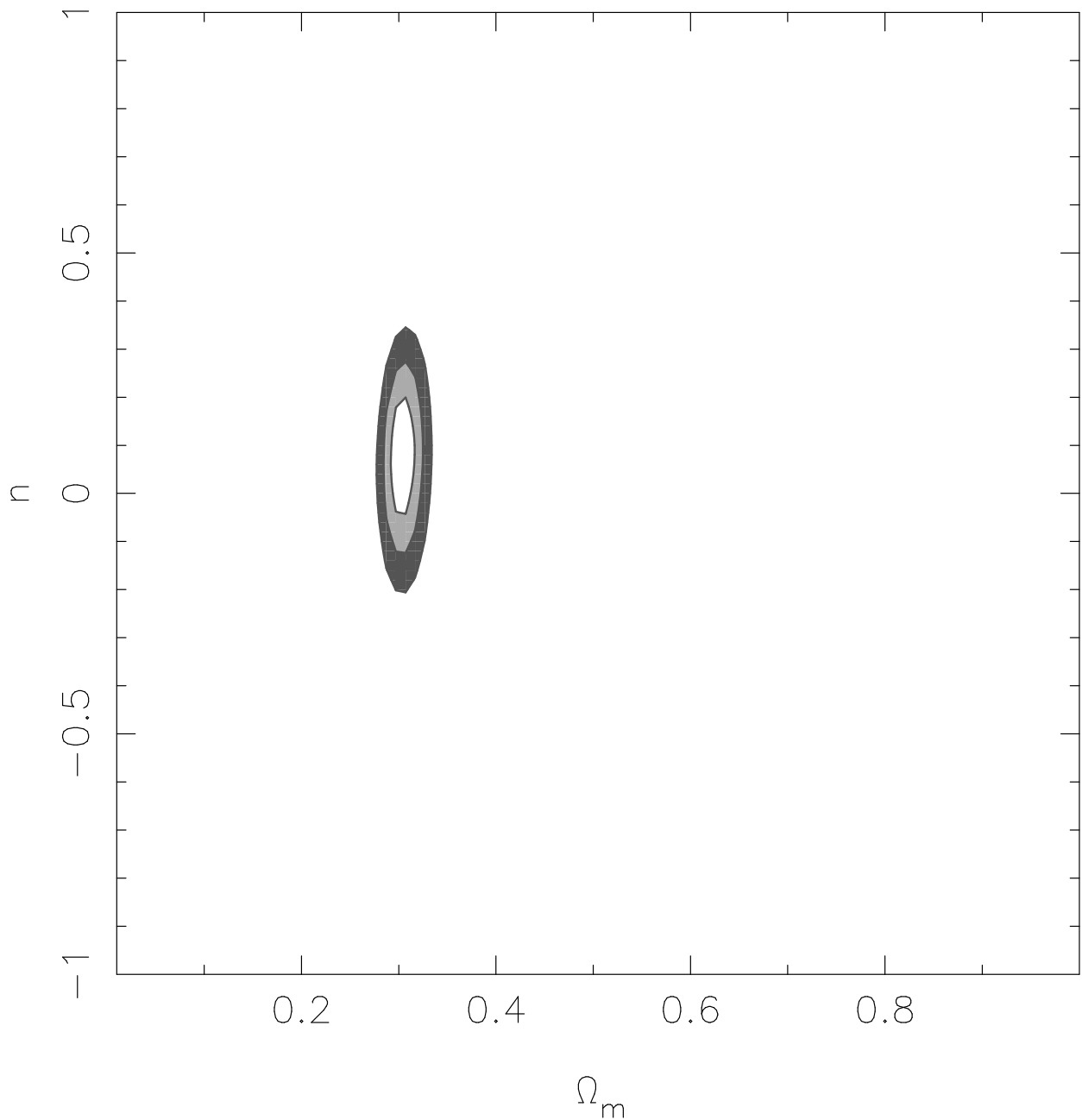}
\figcaption{Confidence region plot of the best fit from a combined analysis
            for the dimensionless coordinate distances to 78 SNeIa and 20 FRIIb
            radio galaxies compiled by Daly and Djorgovski (2003) 
	    and the X-ray gas mass fractions of 9 clusters published by 
	    Allen et al. (2002, 2003).
            The 68.3\%, 95.4\% and 99.7\% confidence levels in the
            $n$--$\Omega_m$ plane are shown in white, white + lower shaded
            and white + lower and darker shaded areas respectively.
	    \label{Fig_cont}
           }
\end{figure}

\end{document}